\documentclass[12pt,preprint]{aastex}

\shorttitle{Equatorially-Asymmetric Magnetized Supernovae}
\shortauthors{Sawai, Kotake, \& Yamada}

\begin{document}

\title{Numerical Simulations of Equatorially-Asymmetric Magnetized
Supernovae: Formation of Magnetars and Their Kicks}

\author{Hidetomo Sawai\altaffilmark{1}, Kei Kotake\altaffilmark{2,3} and
Shoichi Yamada\altaffilmark{1,4}}
\altaffiltext{1}{\textit{Science \& Engineering, Waseda University,
3-4-1 Okubo,
Shinjuku, Tokyo 169-8555, Japan}}
\email{hsawai@heap.phys.waseda.ac.jp}

\altaffiltext{2}{\textit{Division of Theoretical Astronomy, National
    Astronomical Observatory of Japan, 2-21-1 Osawa, Mitaka, Tokyo 181-8588,
    Japan}}

\altaffiltext{3}{\textit{Max-Planck-Institute f\"{u}r Astrophysik,
Karl-Schwarzshild-Str. 1, D-85741, Garching, Germany}}

\altaffiltext{4}{\textit{Advanced Research Institute for Science \&
Engineering, Waseda University, 3-4-1 Okubo,
Shinjuku, Tokyo 169-8555, Japan}}

\begin{abstract}
A series of numerical simulations on magnetorotational core-collapse
 supernovae are carried out. Dipole-like configurations which are offset
 northward are assumed
 for the initially strong magnetic fields together with rapid
 differential rotations. Aims of our study are to
 investigate effects of the offset magnetic field on magnetar kicks
 and on supernova dynamics. Note that we study a regime where
 the proto-neutron star formed after collapse has a large magnetic field
 strength approaching that of a ``magnetar'', a highly
 magnetized slowly rotating neutron star. As a result, equatorially-asymmetric
 explosions occur with a formation of the bipolar jets. We find that the
 jets are fast and light in the north and a slow and heavy in the south
 for in rapid cases while they are fast and heavy in the north and slow
 and light in the south for slow rotation case.
 Resultant magnetar's kick velocities are $\sim 300-1000$~km~s$^{-1}$.  
 We find that the acceleration is mainly due to the magnetic pressure
 while the somewhat weaker magnetic tension works toward the opposite
 direction, which is due to stronger magnetic field in the northern
 hemisphere.
 Noted that observations of magnetar's proper motions are very
 scarce, our results supply a prediction for future observations. Namely,
 magnetars possibly have large kick velocities, several hundred~km~s$^{-1}$, as
 ordinary neutron stars do, and in an extreme case they could have those up
 to 1000~km~s$^{-1}$. In each model, the formed proto-magnetar is slow
 rotator with rotational period of more than 10~ms. It is also found that, in
 rapid rotation 
 models, the final configuration of the magnetic field in the
 proto-magnetar is a collimated dipole-like
 field pinched by the torus of toroidal field lines whereas the
 proto-magnetar produced in the slow 
 rotation model is totally poloidal-field dominant. 
\end{abstract}


\section{Introduction}\label{intro}
 Soft-gamma repeaters (SGRs) and anomalous
 X-ray pulsars (AXPs) are candidates of so-called ``magnetars'', highly
 magnetized slowly rotating neutron stars. This is based on
 the ``magnetar model'', i.e. their burst-like activities, persistent
 X-ray emissions, and large spin-down rates will originate from high 
 magnetic fields, $B \gtrsim 10^{14}-10^{15}$~G
 \citep{dun92,pac92,tho95,tho96}.
 It is notable that such a large magnetic field also may power supernova
 explosions (magnetorotational supernova). Thus, study of magnetorotational 
 supernovae may be important in the context of magnetar formations.  
   
 The first paper concerning magnetorotational supernovae written by
 \citet{leb70} appeared about 30 years before the idea of ``magnetar
 model'' has
 arisen. They found that strong magnetic field combined with rotation
 produced an asymmetric supernova explosion. 
 Only a few studies followed after this \citep{bis75, mul79, ohn83,
 sym84}, reflecting a feeling that
 magnetic fields large enough to drive supernovae were thought to be
 unrealistic. Now that such high magnetic fields are observationally
 supported, the magnetorotational supernova is attracting more
 attention than before. In the last several years some papers have been
 published on the magnetorotational supernovae \citep[e.g.][]{whe02,
 aki03, yam04, kot04, tak04, ard05, kot05, saw05, obe06, 
 nis06, moi06, shi06}. Although, with these studies we have developed
 our understanding of magnetorotational supernovae, many issues still
 remain to be investigated. For example, neutron star's kicks in the
 context of magnetorotational supernovae (magnetar's kicks) have not been studied fully numerically. (see, however, \citet{kot05}: they computed the
 parity-violating effects in strongly magnetized supernova cores and
 expected that neutron star's kicks would result.) In fact, all numerical
 simulations of magnetorotational core-collapse so far assumed equatorial
 symmetry and computed only a
 quarter of a meridional plane of the star.

 Observed neutron stars are known to have generally larger proper
 velocities than that of their progenitor stars \citep{gun79}. Their
 three-dimensional velocities are typically several
 hundred~km~s$^{-1}$ and some of them have a speed higher than
 1000~km~s$^{-1}$ \citep[e.g.][]{cor98,hob05}. The origin of the proper
 velocities is still controversial. Although many researchers
 agree that an asymmetric supernova explosion is a more promising origin
 rather than a binary-system
 disruption by a supernova explosion, several different mechanisms have been
 proposed, which can be categorized mainly into three
 classes \citep{lai03}. 

 The first mechanism is a hydrodynamically-driven
 kick. \citet{sch06} numerically studied hydrodynamic instabilities
 in neutrino-driven supernova explosions and showed that neutron stars
 were accelerated up to about 500 - 1000~km~s$^{-1}$ due to the dominance of $l
 = 1$ mode. In their numerical simulations  neutron stars are substituted
 with gravitating inner boundaries. \citet{fry04} performed
 three-dimensional SPH simulations and asserted that neutrino asymmetries
 decelerated a kicked neutron star significantly, which \citet{sch06}
 argued were negligible. The second mechanism is
 a neutrino-magnetic-field driven kick. \citet{lai98} studied a
 possibility that an asymmetric field topology could produce a large kick
 when combined with neutrino emission. They pointed out problems in
 former similar works and improved on how to deal with microphysics. The
 conclusion which they derived was that at least $\sim 10^{16}$~G difference of
 magnetic field strength between the north and south poles of a
 proto-neutron star is 
 necessary to get 300~km~s$^{-1}$ kick velocity. \citet{arr99}
 investigated parity violation effects on neutron star's kick with again
 improved microphysics. They argued that for the generation of the kick
 velocity of a few
 hundred~km~s$^{-1}$, at least $10^{15}-10^{16}$ G
 of dipole magnetic field is necessary in a proto-neutron star.  
 The third mechanism is an electromagnetically-driven kick, which was
 suggested by \citet{har75}. According to their
 analysis neutron star's kick velocity of several hundreds~km~s$^{-1}$ can
 be generated when the initial rotation period of pulsar is $\sim 1$ ms
 though with the initial rotation period of $\sim 10$ ms the result will
 be two orders of magnitude smaller than the former velocity. This mechanism
 does not depend on the strength of magnetic fields. The main difference
 from the other two mechanisms is a time scale of kick generation. With
 a surface field of $\sim 10^{12}$~G and an initial rotation period of
 $\sim 1$ ms, the acceleration time scale of a pulsar is approximately five
 years.

In this paper we consider an another class mechanism,
``\textit{magnetohydrodynamically-driven kick}''. Consider the
situation that the rapidly rotating core has the dipole-like magnetic
field which is somewhat offset from the center of the core   
prior to collapse. Then, it is expected that a supernova with
equatorially-asymmetric bipolar jet will occur. This
will lead to a kick of the neutron star. Magnetar-class strong magnetic field
is required for this mechanism to work. If this were to a kick mechanism 
of ordinary neutron stars, the magnetic field must decay from 
$\sim10^{15}-10^{16}$~G to $\sim 10^{12}$~G. (This is
also the case for the second mechanism above.) \citet{gol92} estimated
the decay time scales of the magnetic field
in a neutron star due to Ohmic dissipation or ambipolar
diffusion. Their results imply that the magnetar-class magnetic fields
confined in 
neutron stars require at least $\sim 10^6 - 3 \times 10^7$ years to
decay to the strength of typical neutron stars ($\sim 10^{12}$ G). This
is much longer than the ages of not small number of ordinary
pulsars. Hence, we ignore the kicks of ordinary pulsars here, but are
concerned only with the kicks of magnetars.  

  According to the ``\textit{magnetohydrodynamically-driven kick}'' mechanism,
 we numerically study the core-collapse of massive star with off-centered 
 strong diploe-like magnetic fields and rapid rotation.

 The initial magnetic 
 field strength chosen here corresponds to the magnetar-class or larger 
 ($\sim10^{15}-10^{16}$ G) when the core contracts to a typical neutron-star 
 radius. This means that we adopt a so-called fossil origin hypothesis of 
 magnetic fields in magnetars. Although the origin of magnetic fields
 in magnetars are still a mystery, \citet{fer05} asserted that the magnetic
 flux of $\theta$ Orionis C, which is an O star, corresponds to that of
 magnetars and that the magnetism of magnetars could be explained as the
 fossil of progenitors as in the case of white dwarfs
 \citep{wic05}. In
 fact, the magnetic flux of $\theta$ Orionis C is
 $\sim~1\times10^{27}$~G~cm$^2$, which corresponds to $\sim~10^{15}$~G of
 the surface field if a neutron star is formed with the frozen magnetic
 field. There are several other OB stars suggested 
 observationally to possess magnetar-class magnetic fluxes. For example, 
 the estimated magnetic flux of HD191612 \citep{don02} corresponds to the
 surface field strength of $\sim~5\times10^{15}$~G, whereas several
 $\sim~10^{14}$~G of surface magnetic fields are expected for $\zeta$
 Cassiopeiae \citep{nei03a}, $\omega$ Orionis \citep{nei03b}, $\xi$ CMa
 \citep{hub06} and V2052 Ophiuchi \citep{nei03c} if their magnetic fields are
 compressed in the formation of a neutron star. Among them, $\theta$ Orionis C
 and HD191612 are O stars while the others are B stars. On the other hand,
 \citet{tho93} proposed that strong magnetic fields of magnetars originate
 from the convective dynamo process, which requires a rotation period shorter
 than $\sim 30$ ms. However, whether such a process really operates in neutron
 stars or not, is still uncertain. In this paper, we take a scenario that some
 OB stars have large magnetic fluxes and end up with a magnetar (fossil origin
 hypothesis). Although the initial magnetic fields assumed in this study are
 still a little larger than those implied by above observations, we are
 exploring the extreme limit. 

The offset of
the magnetic field in our computation may be supported by numerical 
simulations of fossil magnetic field in A stars and white dwarfs carried
by \citet{bra04}. They found that initially random magnetic
fields evolved into a coherent dipole-like field configuration stabilized by
a torus of twisted magnetic field lines. This magnetic field diffuse
within $\sim 10^9$~years which is much longer than life time of massive
stars. We would like to stress that the resultant magnetic field was
usually somewhat offset from the center of the star. Note that they assumed
fossil origin hypothesis of magnetic field which we also adopt here.   

The rotation period assumed in our rapid rotation models (see
\S~\ref{model}) is an order of magnitude shorter than the result of
\citet{heg05} who calculated the evolution of rotating massive
stars. In spite of the importance of their study, it is not still the
conclusion. The angular momentum distributions at the pre-collapse stage
should be studied further in the future. Our position is that they are
not well-known yet and just consider the extreme case.    

Since observations of magnetar kick velocities are very scarce so far,
our standpoint in this paper is predicting future 
observations of magnetar's proper velocities. Proper motions 
have been measured for only two magnetar candidates (both are AXPs), in
which just upper limits were put; $\lesssim 1400(D/10$ kpc$)$~km~s$^{-1}$ for
4U0142+61  \citep{hul00,woo06} and $\lesssim 2500(D/3$ kpc$)$~km~s$^{-1}$
for 1E2259+586 \citep{oge05}. \citet{gae01} examined the reliability of
AXP/SGR association with supernova remnants (SNRs). The associations
which they considered probable enable us to estimate velocities of
AXPs and SGRs from their positions in the SNRs: $<$ 500~km~s$^{-1}$ for AXP
1E 1841-045, $<$ 400~km~s$^{-1}$ for AXP 1E
2259+586, and $<$ 500~km~s$^{-1}$ for AX J1845-0258 (AXP candidate)
\citep[][and reference their in]{gae01}. \citet{cli82} evaluated the
velocity of SGR 0526-66 assuming that it is associated
with SNR N49 and derived
$v_{kick} \sim 400-1400$ km~s$^{-1}$. \citet{cor99} derived $\sim
800-1400$~km~s$^{-1}$ velocity of SGR 1627-41 from the association with
SNR G337.0-0.1. However, \citet{gae01} insisted
that these two associations between the SGRs and the SNRs are
less convincing. 

Recently, we reported the results of
numerical study that configurations of magnetic fields affect supernova
dynamics significantly \citep{saw05}. The effects of the offset dipole
field on the supernova explosion is another subject which we are
interested in this paper. We particularly focus on how dynamics and the
explosion energy are altered and that north-south difference of the
magnetic and rotational energies. 

The rest of this paper organized as follows. We describe numerical
models such as basic equations, a equation of state, magnetic field and
rotation, and the definition for proto-magnetar in $\S$~2. In $\S$~3,
the some numerical results including dynamics and kick velocities are
shown. In $\S$~4, we give some discussions and conclude the paper. 

\section{Models}\label{model}
We numerically simulate the core-collapse of magnetized massive stars
with the numerical code ZEUS-2D developed by \citet{sto92}.
A $1.5M_\odot$ core of $15 M_\odot$ star
provided by \citet{woo95} is chosen as a pre-collapse model, which gives
spherically
symmetric profiles of the density and specific internal energy. 
Magnetic field and rotation are just added to the core by hand (see \S
\ref{magrot}). 

\subsection{Basic Equations}
The following ideal MHD equations are solved:
\begin{eqnarray}
  \frac{D\rho}{Dt}+\rho\nabla\cdot\mbox{\boldmath $v$}=0\label{mac},\\
  \nonumber\\
  \rho\frac{D\mbox{\boldmath $v$}}{Dt}=-\nabla p-\rho\nabla\Phi+\frac{1}{4\pi}
   (\nabla\times\mbox{\boldmath $B$})\times\mbox{\boldmath $B$}\label{moc},\\
  \nonumber\\
  \rho\frac{D}{Dt}(\frac{e}{\rho})=-p\nabla\cdot
   \mbox{\boldmath $v$}\label{enc},\\
  \nonumber\\ 
  \frac{\partial\mbox{\boldmath $B$}}{\partial t}=\nabla\times
   (\mbox{\boldmath $v$}\times\mbox{\boldmath $B$}),\label{far}
\end{eqnarray}
where $\rho$, \mbox{\boldmath $v$}, $p$, $e$, $\Phi$, \mbox{\boldmath
$B$} are the density,
velocity, internal energy density, gravitational potential,
and magnetic field, respectively. We denote
the Lagrangian derivative as \mbox{$\frac{D}{Dt}$}.

A computational region is a half of the meridional plane with axisymmetric
assumption. $200(r) \times 60(\theta)$ grid points extending to 2000~km
are used initially. After central density reaches $10^{12}$ g
cm$^{-3}$, we change the radial mesh resolution and a radius of outer
boundary into $300(r)$ and 1500~km, respectively. The radial grid spacing are
not uniform with a finer resolution for smaller radii, while the angular
grid points are put uniformly. 

\subsection{Equation of State}\label{eos}   
As in the previous papers by \citet{yam04} and \citet{saw05}, we use a
parametric EOS
which was first introduced by \citet{tak84}. Although we
drastically simplify complicated microphysics such as the
neutrino transport, it is not a serious concern since our computation is
done only on the time scale of prompt explosion, during which the neutrino
transport is less important.

The parametric EOS we employed in this paper is as follows;
\begin{eqnarray}         
 p_{tot}=p_c(\rho)+p_t(\rho,e_t),\label{eosm}\\
 p_c(\rho)=K\rho^\Gamma,\label{eosc}\\ 
 p_t(\rho,e_t)=(\gamma_t-1)\rho\epsilon_t\label{eost}.
\end{eqnarray}
The pressure consists of two parts, the cold part ($p_c$) and the thermal
part ($p_t$). The thermal part is a function of the density and the
specific thermal energy, $\epsilon_t$, in which $\gamma_t$ is the
parameter called the thermal stiffness. The cold part is a function
of the density alone where the two constants, $K$ and $\Gamma$, take
account of the effect of the degeneracy of
leptons and the nuclear force. We choose the same values for $\gamma_t$,
$K$ and $\Gamma$ as in \citet{saw05}, so that a numerical simulation without
either magnetic filed or rotation fails to explode as in recent realistic
simulations. Readers are referred to the paper by \citet{saw05} for more details about this EOS.

\subsection{Magnetic field and Rotation}\label{magrot}
The initial magnetic field is assumed to have a dipole-like structure produced
by the sum of the line currents
which are uniformly distributed in the spherical region within a 1500~km
radius. The equatorially-asymmetric dipole magnetic field is obtained by
displacing equatorially-symmetric field along the 
rotation axis. Then the initial magnetic filed in
the cylindrical coordinate is given by
\begin{eqnarray}
B_{\varpi}(\varpi, z) = \frac{2J}{c} \sum_{\varpi_J} \sum_{z_J}\frac{z^{'}}{\varpi \sqrt{(\varpi_J + \varpi)^2 + z^{'}}}\Bigg[-K + \frac{\varpi_J^2 + \varpi^2 + z^{'2}}{(\varpi_J - \varpi)^2 + z^{'2}} E \Bigg],\label{emag1}\\
B_z (\varpi, z)= \frac{2J}{c} \sum_{\varpi_J} \sum_{z_J}\frac{1}{\sqrt{(\varpi_J + \varpi)^2 + z^{'}}}\Bigg[ K + \frac{\varpi_J^2 - \varpi^2 - z^{'2}}{(\varpi_J - \varpi)^2 + z^{'2}} E \Bigg],\label{emag2}\\
z^{'} = z - z_J - z_{off},\label{emag3}
\end{eqnarray}
where $R$ is a distance from the center, and $(\varpi_J, z_J)$ is the
mesh point where the line currents
exist. $J$, $K$, $E$, $c$, and $z_{off}$ are the current, complete
elliptical integral of the first kind, complete elliptical integral of
the second kind, the speed of light, and the degree of offset,
respectively. 
We compute with changing the magnetic field strength and
degree of offset (see Table~\ref{tmodels}). Note that, if the result of
\citet{bra04} is adopted to the iron core of a 2000~km radius, the
degree of displacement is several hundred~km. Fig.~\ref{fbfield} shows
the initial magnetic field configuration of models MR3 and SR10 (for the
name of models, see below).
 
The initial angular velocity in our simulations is assumed to have a
differentially rotating distribution,
\begin{equation}
\Omega(r)=\Omega_0\frac{r_0^2}{r_0^2+r^2},
\label{dif}
\end{equation} 
where $\Omega_0$ and $R_0$ are constants. With this distribution,
rotation is faster at small radii. In all models,
$R_0 = 1000 $~km is chosen, with which the initial differential rotation is
rather mild. We compute the models with rapid
rotation and slow rotation cases (see Table~\ref{tmodels}). The rapid
and slow rotation correspond to a neutron star rotating with the period
of $\sim 1$~ms and $\sim 10$~ms, respectively, if the angular
momentum conservation is assumed.

In Table \ref{tmodels}, six models computed are summarized. The
name of each
model consists of three parts, two characters and a number. The first
character denotes a strength of magnetic field; ``Moderate'' and
``Strong''. The second character represents a rotation period, ``Slow''
and ``Rapid''. The attached number stands for a degree of
magnetic-field displacement.
For example, model MR3 has a moderate magnetic field, rapid rotation and
the displacement of 300~km. 

\subsection{Definitions of a Proto-Magnetar and its
  Velocity}\label{nsdef} 
We define a proto-magnetar (or a proto-neutron star) in our analysis
as a region which has a mass of $1.2 M_\odot$. The mass of each fluid
element is summed up descending order of the density. This is almost the 
maximal mass of fluid which remains through the simulations in the numerically
active region and does not go out of the outer boundary.  
Neutron stars in binary systems typically have the mass of around
$1.4M_{\odot}$ \citep[e.g.][]{tay82,tho99}. However, the mass of
magnetar candidates is 
unknown at present. \citet{sch06} discussed that
resultant mass of neutron star decreases as explosion time scale becomes
short. If explosion occurs in a prompt manner, a $1.2 M_\odot$ magnetar
may not be so bad. 

The velocity of the proto-magnetar is defined as follows,
\begin{eqnarray}         
v_{NS}(t) = \int_{t_0}^{t} \frac{F(t')}{M_{NS}} \mathrm{d}t', \label{evnsdef1}\\
F(t) = \oint_{S_{NS}} ( - P \mbox{\boldmath $n$} - \frac{B^2}{8 \pi} \mbox{\boldmath $n$} + \frac{B_z}{4 \pi} \mbox{\boldmath $B$} ) \cdot \mbox{\boldmath $n_z$} \mathrm{d} S - \int_{V_{NS}} \rho \nabla \Phi_{outer} \mathrm{d}V, \label{evnsdef2}
\end{eqnarray}
where, \mbox{\boldmath $n$}, \mbox{\boldmath $n_z$}, and
$\Phi_{outer}$ represent a local unit
vector normal to each surface element on a proto-magnetar, a unit
vector parallel to the rotation axis, and the gravitational potential
attributed 
to mass distribution in the outer layer of the proto-magnetar,
respectively. $t_0$ is the time when the central density reaches 
$1.0 \times10^{12}$~g cm$^{-3}$, before which the acceleration of the
proto-magnetar is negligible.

The velocity of a magnetar also could be defined in principle as the
 center of mass velocity. However we do not adopt this definition here
 because it leads to enormous numerical errors in calculating the volume
 integral of momenta over the region of the proto-magnetar. 
 The velocity at each grid point is obtained by the time-integration of the 
 volume forces (pressure gradient, Lorentz force and gravity) exerted there
 and the advection term. Hence, the volume integral of the momentum over 
 the proto-magnetar is equal to the time integration of the volume integral of
 these forces and the advection term over the proto-magnetar. 
 The difficulty in the numerical estimation of the proto-magnetar's kick
 velocity lies in the volume integral of the forces and the advection term. In
 fact, the error in calculating the proto-magnetar's kick velocity in this way
 comes not only from the surface of the proto-magnetar but also from its
 entire volume, and the problem is that the values of the integrand at the
 center are greater by a few orders of magnitude than those at the surface of
 the proto-magnetar and, as a result, errors of a few percent around the
 central region of the proto-magnetar correspond to errors of more than 100
 percent around the surface. On the other hand, if we calculate the
 proto-magnetar's kick velocity by time-integrating the surface forces, the
 resultant numerical errors originate from the local errors in the forces on
 the proto-magnetar surface and their magnitude will be comparable to the
 typical error, namely a few percent. This corresponds to an error of a few
 10 km/s in the proto-magnetar's kick velocity in model MR3, for example.
 It should be noted that the employment of a conservative scheme does not
 guarantte the correct spatial distribution of momentum. Moreover, since the
 self-gravity cannot be written in a conservative form, the strict
 conservation of the total momentum cannot be expected after its inclusion.

\section{Results}\label{result}
We show the numerical results of our computation in this section.
The explosion dynamics, explosion energies, final states of
magnetic field and rotation, and magnetar kicks are given separately.

\subsection{Dynamics}\label{dyn}
We first describe the dynamics of each model. The important parameters
for all models are summarized in Table 2 and~\ref{texp}. 

Model MR0 has the same
initial parameters as model C10 of \citet{saw05} except for the magnetic
field configuration. The magnetic field of C10 was parallel
to the rotation axis and its strength was stronger near the rotation
axis while the magnetic field of model MR0 is dipole-like with larger
strength near the center. Namely, models MR0 is 
a more ``centrally'' concentrated counterpart of model C10. The
dynamical evolution of model MR10 is described as follows.
Model MR0 first bounces at 142.7 ms after the onset of collapse when the
central density reaches $3.75 \times 10^{14}$~g~cm$^{-3}$. However the
first shock wave is generated not by the nuclear force but by the magnetic
force  0.2 ms prior to the nuclear bounce around a radius of 30~km near the
rotation axis, which was also the case for model C10 in \citet{saw05}. 
Indeed the magnetic field around the region is strong enough by the time
of bounce 
due to the compression and field wrapping that the magnetic pressure becomes
much larger than the matter pressure. Then, the magnetic-force-dominant regions
gradually spread outward and shock waves propagate through the core
accompanied by the formation of a bipolar jet. This occurs by
transferring the magnetic energy to the kinetic
energy. As shown in left panel of Fig.~\ref{fengy} and its inset, the
toroidal energy is produced by the 
rotational energy especially around the time of the first bounce ($\sim
143$~ms) and the subsequent second bounce ($\sim 147$~ms). After the
amplifications, the toroidal magnetic energy clearly 
decrease as the kinetic energy of matters whose radial velocity is
positive (the outward kinetic energy) increases. The toroidal energy is
not only consumed to launch the jets but also continues to be produced by
the rotational energy during this phase. This is why the rate of the outward
kinetic energy increase is 
larger than that of the toroidal energy decrease. The poloidal field
energy also will be transferred into the kinetic energy although the
total poloidal magnetic energy itself increases with time as the matter
falls. The outer most shock front reaches a 
radius of 1500~km, 20~ms after the bounce (see top left panel of
Fig.~\ref{fvbeta}). The 
explosion energy at that time is  $4.4\times 10^{51}$~erg, which is
almost twice as large as the most explosive model C10 in
\citet{saw05}. This implies that centrally concentrated magnetic fields
tends to explode supernovae more powerfully. 

Offset magnetic field models with rapid rotation, MR3, SR3 and SR10,
produce equatorially-asymmetric explosions with the faster 
jet northward, in which the degree of asymmetry is the most 
remarkable for model SR10. (see Fig.~\ref{fvbeta}).  
In these models, shock waves are
generated also prior to the nuclear bounce due to the magnetic force,
which occurs first in the northern hemisphere 
and then in the southern hemisphere with $\sim 0.1-0.6$~ms delay. Then
the shocks propagate through the core with larger speed northward.
The magnetic energy is more dominant in the
northern hemisphere than in the southern hemisphere almost always until
the shock reaches the surface of the core although the toroidal part is
larger in the southern hemisphere (Fig.~\ref{fengmag}). This is why
the jet and shock propagation are faster in the north.

We show the reason why the toroidal magnetic energy is larger in the
southern hemisphere.
The time evolutions of the rotational energies in Fig.~\ref{fengmag}
show that matter collapses more deeply in the northern 
hemisphere around the first bounce for models MR3 and SR3\footnote{The
two bumps in the rotational energy evolution in each panel of
Fig.~\ref{fengmag} roughly show how deeply the core collapses,
reflecting the gravitational energy evolution.}. This is because
the stronger magnetic fields in the northern hemisphere extract the
rotational energy more efficiently, which results in that the
centrifugal force reduces and the matter falls favorably. On the other
hand, the same interpretation is not valid for model SR10, in which the
collapse is deeper in the southern hemisphere in turn around the first
bounce. It is considered that magnetic field not only
encourages the collapse by 
extracting the rotational energy, but also discourages it. The latter
effect occurs simply because the
magnetic force preclude the matter from falling where it is dominant,
i.e. around the rotation axis with a radius more than $\sim 100$~km for
the rapid rotation models.
Note that, for every model computed here, there always locally exist the region
where the magnetic pressure dominates the matter
pressure during collapse, and so this effect is significant in spite of the
fact that the total magnetic energy is much smaller than the rotational
energy. This effect seems to work well in model SR10. It is
likely that too strong magnetic field is disadvantageous for the infall
of the matter. For the
second bounce in each model, the magnetic field rather
works to decrease the infall rate which leads to deeper collapse in the
southern hemisphere since the magnetic field grows larger than that at the
first bounce. As a result,
the rotational energy becomes larger in the
northern hemisphere after the second bounce, and thus the toroidal
magnetic energy also becomes larger. 
 
Table~\ref{texp} shows that the mass ejection in the
southern hemisphere is larger than that in the northern
hemisphere for model MR3\footnote{The ejected mass is defined as the sum
of the mass of fluid elements with the positive total energy. With this
definition, a part of mass in the proto-magnetar is also summed.}. 
This is explained by the effect mentioned above, i.e. the magnetic force
prevents the matter infall. An amount of ejected mass is determined by
how much matter is accumulated around the outside of the boundary between
the inner and outer cores when the shock waves are generated. In fact,
it is the outer core where the braking of infall by the magnetic
forces works. Owing to stronger magnetic field, the mass infall in the
outer core is more effectively hampered
in the northern hemisphere although the
larger gravitational energy is extracted in the northern hemisphere
around the first bounce for models MR3 and SR3 and so at least the inner
core collapses well in this half. As a
result, mass eruption is smaller in the northern hemisphere. 
According to \citet{yam94}, the amount of available gravitational energy
is sensitive to the ejected mass for the prompt explosion with pure
rotation. However it will not be said for magnetorotational case. 

In summary, we find that the jets are right and fast in the northern
hemisphere and heavy and slow in the southern hemisphere for the rapid
rotation cases.

The slow rotation models, MS3 and SS3, produce an order of magnitude
weaker explosions than rapid rotation models. The expelled masses are
also smaller by an order of magnitude (see Table~\ref{texp}).   
In each of the two models, the magnetic field plays a dynamical role
only just behind the shock fronts as they expands
outward. However, unlike in the rapid rotation cases, the toroidal
magnetic field plays almost no role to power the shock but probably the
poloidal field single-handedly does. In fact the slow rotation can not yield
strong toroidal fields whose energy is enough to help explosion (see
Fig.~\ref{fengy}). The farthest shock front reaches a radius of 1500~km
about 70 ms after the bounce, which is 3.5 times longer than for model MR3. 
Contrary to the rapid rotation models, the slow rotation models expel
more mass in the northern hemisphere (see Table~\ref{texp}). This is
because the magnetic 
pressure during bounce is not strong enough to hamper the collapse (see
right panel of Fig.~\ref{fengy}) and the magnetic field just acts to extract
the rotational energy. Thus for the slow rotation models, the
jets are heavy and fast in the northern hemisphere and right and slow in
the southern hemisphere.

\subsection{The Explosion Energy}\label{seng}
The explosion energies for all model is given in Table~\ref{texp}.
They are estimated by the sum of the energies over the
fluid elements with the positive total energy when the shock front
reaches a radius of 1500~km. It can be seen that for the rapid rotation
models the offset dipole field
weakens the explosion in comparison of MR0 with MR3 and of SR3 with
SR10. For model MR3, the explosion energy in the northern hemisphere is
somewhat smaller than that of model MR0 while the explosion energy in
the southern hemisphere is just slightly larger than that of model MR0,
which makes model MR3 weaker exploder. These features are just reflected
by the amounts of the ejected mass, i.e. the larger mass ejected, the
stronger the explosion becomes (see Table~\ref{texp}). Namely, in this case, the
explosion energy is also controlled by the degree of infall in the outer core as
the amount of the ejected mass. For the comparison between models SR3
and SR10, the situation is different. Although the expelled mass is
larger for model SR10, the explosion energy is larger for model SR3.
This is because the stored magnetic energy after the bounce for model
SR3 is much larger than that for model SR10 both in the northern and southern
hemispheres (see Fig.~\ref{fengmag}). For model SR10, the magnetic field
is very sparse in the southern hemisphere from the beginning. 
Meanwhile, in the northern hemisphere, in spite of initially very strong
magnetic field it can not be amplified so much since the increase of
rotational energy due to collapse is very small (see \S~\ref{dyn} ). 
Not same as the rapid rotation models, the explosion energy is larger in
the northern hemisphere for the slow rotation models. This is simply
because mass ejection is larger in the north (see \S~\ref{dyn}).    
After all, what we find here is that the magnetic fields which induce the
explosions also work against the energetic explosions.    

\subsection{Formation of A Proto-Magnetar}
The proto-magnetars which we call in this paper were defined in
\S~\ref{nsdef}. They have much different shapes
between the rapid rotation models and slow rotation models. In upper two panels
of Fig.~\ref{fmagrot}, the final shape of the proto-magnetars for 
model MR3 and MS3 can be seen. For model MR3, this corresponds to the
region whose density is more than $2.0 \times 10^{7}$~g~cm$^{-3}$ and the
proto-magnetar has a butterfly-like shape in the meridional plane,
occupying large fraction in the computational domain. On
the other hand, the proto-magnetar of model MS3 whose critical density
is $7.8 \times 10^{9}$~g~cm$^{-3}$, has only a radius of $\sim 100$~km
and its shape is a oblate ellipsoid.

Final rotation-period profiles for models MR3 and MS3 are also displayed
in upper two panels of Fig.~\ref{fmagrot}. For model MR3 the constant
$\Omega$ surface in the proto-magnetar is rather cylindrical than
spherical because the matter is expelled strongly toward both poles.
There exist a
slowly-rotating torus with a $\sim 100$~km radius in the inner region of
the proto-magnetar. Starting from the surface of the torus, narrow zones
where the angular velocities are also small run along the inner line of
the wings of the ``butterfly''.
In the proto-magnetar, there is almost no region whose rotation period
is smaller than 10~ms. In model MS3, the
angular momentum distribution is disarray especially at small radii
though at large distances from the rotation axis it is marginally
cylindrical. All fluid
elements rotate with a rotation period of about a few seconds in the
proto-magnetar.  

The magnetic field distribution at the end of
simulations is shown in lower panels of Fig.~\ref{fmagrot}. In model MR3,
a region where the poloidal field is strong spindles along the rotation
axis to the surface of the core due to the formation of the strong bipolar
jet. The poloidal field has a collimated dipole-like shape around the
rotation axis and no
longer much off-centered as before while the toroidal field is somewhat
equatorially-asymmetric. Around the foot of the jets and the inner
part of the proto-magnetar, the toroidal fields are very weak since their
energy is transferred into the kinetic energy of jets. On the other
hand, in the outer part of the proto-magnetar (the wings of the
``butterfly), the toroidal field is
dominant in most area. The global magnetic field structure is roughly such that
the dipole magnetic field pinched by the toroidal magnetic field lines,
the region with more than $\sim 10^{13}$ G regarded.
How this magnetic fields
will evolve during cooling of the proto-magnetar to a magnetar of $\sim
10$~km radius
is of another interest which should be investigated elsewhere in the
future. In model MS3, areas where the toroidal field is comparable to
the poloidal field are very small. There is a cylinder at the center of the
core with
$\sim 1400$~km length and $\sim 400$~km radius (a blue-colored cavity in
Fig.~\ref{fmagrot}) where is totally poloidal-field
dominant. The proto-magnetar is in the most inner part of this cylinder,
in which the toroidal 
field is especially weak. The poloidal field is also non-offset
dipole-like but is not as collimated as in the case of model MR3 owing
to weakly ejected mass.  

\subsection{Kick Velocity of a Nascent Magnetar}
We show, in Fig.~\ref{fvns}, the time evolution of velocities of the
proto-magnetars for all models. Models MR3, MS3, SR3 and SS3 produce kick
velocities of around $350-500$~km s$^{-1}$ while model SR10 yields a
velocity of more than 1000~km s$^{-1}$. Note that in the
equatorially-symmetric model MR0, the velocity of the proto-magnetar is less
than 70~km s$^{-1}$, which are due to un-removable numerical noises. 
The final velocities are estimated at 240 ms
from the beginning of computations. 

In model MR3, the proto-magnetar is
kicked southward and finally reaches a kick velocity of
512~km~s$^{-1}$. As shown in  
Fig~\ref{fans}, the driving force is the magnetic pressure. Since the
magnetic pressure generally pushes the matter and the magnetic energy is
superior in the northern hemisphere (see
Fig.~\ref{fengmag}\footnote{This figure 
does not shows the magnetic energies just around the proto-magnetar but
those integrated over the northern or southern hemisphere. Hence, it
gives just a rough indication here.}), the
proto-magnetar is pushed southward. Meanwhile, the magnetic
tension which generally pulls the matter works northward again with the help
of stronger magnetic field in the northern hemisphere although this is
somewhat weaker than the magnetic 
pressure. Now we discuss what determines which of these two stresses
becomes dominant. The z-component of the force owing to magnetic
pressure and tension of the poloidal field in cylindrical coordinates are  
\begin{eqnarray}         
F_{mp,z} = -\frac{1}{8\pi}\int_{S}B^2 dS_z = -\frac{1}{8\pi}\int_{S}B^2 \cos\theta_2 dS,\label{efmp}\\
F_{mt,z} = \frac{1}{4\pi}\int_{S}B_z \mbox{\boldmath $B$}\cdot d\mbox{\boldmath $S$} = \frac{1}{4\pi}\int_{S} B^2 \cos\theta_1 \cos(\theta_1 - \theta_2) dS\label{efmt},
\end{eqnarray}
where, $B^2=B_z^2+B_R^2$, and $\theta_1$ and $\theta_2$ is the angles of
the magnetic field and 
surface element vector measured from the rotation axis, respectively. We
consider here in the range $\theta_2-90^\circ \le \theta_1 \le \theta_2 +
90^\circ$ and $-90^\circ \le \theta_2 \le 90^\circ$, i.e. the magnetic
field is outward on the surface in the northern hemisphere. 
A contribution by the toroidal field is omitted for simplicity for the
moment. Where the magnetic tension is locally dominant, the following
relation works out from Eqs.~(\ref{efmp}) and ~(\ref{efmt}), 
\begin{eqnarray}         
2B^2 \cos\theta_1 \cos(\theta_1 - \theta_2) - B^2 \cos \theta_2 = \cos(2\theta_1 - \theta_2) \ge 0,
\end{eqnarray}
and then $\theta_1$ satisfies
\begin{eqnarray}         
\frac{\theta_2}{2}-45^\circ \le \theta_1 \le \frac{\theta_2}{2}+45^\circ,
\end{eqnarray}
(see Fig.~\ref{fmagtp}). Regarding the northern hemisphere of model MR3
at 240 ms, Fig.~\ref{fmagrot} indicates that $\theta_2$ is often less than
-~60$^\circ$, where the magnetic force strongly works. Hence, the magnetic 
tension is dominant when $-75^\circ \lesssim \theta_1 \lesssim
15^\circ$. If the effect of the toroidal field is taken into account, the
upper critical angle become somewhat larger because the toroidal field
is considered to be large in the southern hemisphere (see
\S~\ref{dyn}). On the other hand, the left-bottom panel of Fig.~\ref{fmagtp}
shows that $\theta_1$ is adequately more than $15^\circ$ almost
everywhere. Therefore the magnetic pressure is dominant
there and it becomes a leading force to accelerate the proto-magnetar. The same discussion can be done on the southern hemisphere. In fact in
model MR3 (and also in models SR3 and SR10), such a situation is common for
almost all time in the computation.  

Around $150-160$~ms in model MR3, the magnetic acceleration
decreases, which causes kicking-back of the proto-magnetar
northward. This decrease is likely to be yielded by the predominant
amplification of the toroidal field in the southern hemisphere around that time
(see Fig.~\ref{fengmag}), which makes the magnetic pressure work
northward. After kicking-back, the magnetic stress comes
back to life with even greater strength than before. 
This is because the difference of the magnetic stress
between the north and south surfaces becomes larger although the stresses
themselves are weaker than before. Then the
acceleration gradually decreases as the north-south difference in magnetic
stress diminishes. At the end of the computation, the acceleration is
quite small, and so the proto-magnetar no longer will be accelerated
significantly. Compared with the magnetic stress, the
pressure and gravity are less important for the acceleration of the
proto-magnetar except for the time of kicking-back. 
After the northward-shock
front passes the north surface of the proto-magnetar in first around
150~ms, the pressure is larger in the vicinity of the north
surface. However, this is until the southward shock front passes the south
surface of the proto-magnetar with a few~ms delay. After this, the pressure
in the vicinity of the south surface is always larger because there is
more matter or the heat energy. The gravity pulls the
proto-magnetar to the direction where there is more matter. However,
after 162~ms, a part of matter is ejected out of the numerical domain, but we
neglect the gravity from the expelled matter. Hence, the value of the
gravitational acceleration in Fig.~\ref{fans} is not valid after this
time. Nevertheless,
keeping in mind that the gravitational acceleration is comparable to the
acceleration owing to the pressure, the effect of the gravity on the
kick velocity is no significance either way. The same can be said for
models SR3 and SR10. For models MS3 and SS3, since the ejected mass is
$\sim 0.03 M_\odot$, the gravitational
acceleration changes at most $\sim 10$ \% even if they are taken into account.

In models SR3 and SR10, the time evolution of magnetar velocities and
the acceleration mechanisms
are similar to that of model MR3. In model SR3, however, the magnetic
stress which has accelerated the proto-magnetar southward turns to work
oppositely around
210 ms and the proto-magnetar starts to be decelerated. This is probably
because 
the magnetic field becomes locally stronger around the south surface of
the proto-magnetar. At the end of the
simulation, the magnetic stress still has a positive value. Thus, although
the proto-magnetar gets the final velocity of 353~km s$^{-1}$, this
will be decreased later in some degree. Model
SR10 produces the highest velocity of proto-magnetar among all
models. In spite of 
the smaller stored magnetic energy than that in model SR3, a considerable
north-south difference of them can produce a larger
accelerations. In this model, kicking-back occurs due not to reduction of
magnetic stress but to the pressure and gravity. At the end of
simulation the acceleration is still 
substantial. It will be negligible around 300 ms if linearly
extrapolated. The velocity of the proto-magnetar around that time will
be $\sim 1500-1600$~km~s$^{-1}$. 

The slow rotation models, MS3 and SS3, show more complex time evolution of
the proto-magnetar acceleration (see Fig~\ref{fans}). In each of these cases,
the magnetic pressure alone is not always the leading driving force, but
the magnetic 
tension, matter pressure or gravity also play important roles. This is
due to a low ratio of the magnetic pressure to
matter pressure in the vicinity of the proto-magnetar. Although these
models explode weakly, the final  
magnetar velocity are not differ significantly from those for the more
strong explosion models. This is because the kinetic energy of a
proto-magnetar whose
velocity is $\sim$ a few 100~km~s$^{-1}$ is just $\sim 10^{48}$~erg which is
even smaller than the explosion energies of model MS3 and SS3.
The time evolution of acceleration still varies 
violently even around the end of the computation. In order to get
reliable final velocities, longer numerical simulations are necessary.

\subsubsection{Uncertainties from the definition of A Proto-Magnetar}
 As noted in \S~\ref{nsdef}, we assume a proto-magnetar as region containing
 1.2~M$_\odot$, which are calculated by summing up the mass of each fluid
 element in descending order of the density. However, it is admittedly
 highly uncertain at 240~ms which part of the gas is finally settled to be a
 proto-magnetar. In fact, this will not be known for sure until computations
 are done up to the time, by which a fall back of matter has been finished.
 According to Zhang et al. (2007),
 this time is estimated to be about $10^6$~s, which is far longer
 than our computational time, 240~ms, and is, unfortunately, too long to
 dynamically follow with numerical simulations like ours. What we can do and
 we should do instead is to make clear the uncertainty of the kick
 velocities associated with the definition of a proto-magnetar. For this
 purpose, we have computed another model, which covers a greater portion of
 the star, and we have also calculated the kick velocities with several
 alternative definitions of a proto-magnetar, taking model MR3 as an example.

   We first investigate the uncertainties coming from the proto-magnetar
 mass. Five different masses of proto-magnetars, 1.2, 1.23, 1.26, 1.3,
 1.35~M$_\odot$, are tried. We take here the maximal mass to be
 1.35~M$_\odot$ anticipating that a proto-magnetar produced by a magnetic
 prompt explosion will be less massive than 1.4~M$_\odot$, the canonical value
 for the ordinary radio pulsars. We then introduce another criterion for a
 proto-magnetar, by which fluid elements are summed up in the descending order
 of "pressure" instead of density. Several masses of proto-magnetar are
 tried again with this "pressure criterion". In the study of proto-magnetars
 more massive than 1.2~M$_\odot$, a larger portion of the progenitor star
 needs to be computed. We thus run another numerical simulation, which
 computes the evolution of the inner 2.0~M$_\odot$ of the 15~M$_\odot$
 progenitor by \citet{woo95} for the same parameter values of $|E_m/W|$,
$|T/W|$, and z$_{off}$ as in model MR3, and obtain the kick velocities
 of 10 different proto-magnetars. We give each model the name as
 MR3env[D or P] [proto-magnetar's mass], where ``D'' and ``P'' denote the
 density- and pressure-criterion, respectively. For example, a model for
 1.23~M$_\odot$ proto-magnetar determined by the pressure-criterion is
 referred to as MR3envP1.23.

   The results are shown in Fig. 9, which illustrates the time evolution of
 the velocity for each proto-magnetar. We find that the difference in
 the criterion does not produce a significant variation in the kick
 velocities. On the other hand, it is found that the difference in the
 mass of proto-magnetar results in a noticeable variation in the kick
 velocities. In both series, MR3envD and MR3envP, the velocity tends to
 decrease with increasing mass. This is because a proto-magnetar with a higher
 mass has a larger radius and the magnetic field on the surface becomes
 weaker, which then leads to a smaller north-south difference in the magnetic
 forces. Note, however, that the most massive proto-magnetar models in our
 analysis, MR3envD1.35 and MR3envP1.35, still have kick velocities of
 $\sim$~200~km~s$^{-1}$.

   The kick velocities obtained with several different definitions of a 
 proto-magnetar cover a range of $\sim~200-500$~km~s$^{-1}$. We thus consider
 that the kick velocities we have procure in this study have uncertainty of
 this degree. We can thus claim that a proto-magnetar is possibly accelerated
 up to several hundred~km~s$^{-1}$ by the magnetohydrodynamic process
 considered in this paper.

\section{Discussion and Conclusion}
We have done series of numerical simulations of core-collapse supernovae
with strong offset-dipole magnetic fields (offset northward) and rapid
rotations. Main 
purposes of our study was investigating how dynamics and magnetar
kicks behave with such off-centered magnetic fields. We finally found that:

1. Equatorially-asymmetric explosions occur for all models with a
 formation of bipolar jets. Jets are fast-light in the north and
 slow-heavy in the south for the rapid rotation models while they are
 fast-heavy in the north and slow-light in the south for the slow
 rotation models. 

2. Off-centered magnetic field weaken the explosion for the rapid
 rotation case. In this case, the explosion energy is larger in
 the south since less matter is ejected or less magnetic energy is
 stored in the north. On the other hand, the explosion energy is larger
 in the north for the slow rotation models.

3. The formed proto-magnetars are slow rotator with the rotational periods of
 more than 10~ms. The final magnetic field around the proto-magnetar has
 a collimated dipole-like configuration pinched by the toroidal field
 lines for the rapid rotation models whereas the proto-magnetar formed
 in the slow rotation models is totally poloidal-field dominant. 

4. If the initial magnetic field is stronger in the northern hemisphere,
 proto-magnetars are kicked southward with the velocities of several
 hundred~km~s$^{-1}$, or in extreme case $\sim$~1000~km~s$^{-1}$.   
 In most cases, they are accelerated mainly by the magnetic pressure
 while the somewhat weaker magnetic tension works the opposite
 direction, which is due to stronger magnetic field in the northern
 hemisphere. 

Form the results of computations, we predict that magnetars also
possibly have large velocities as ordinary pulsars do and in some
extreme cases they could have $\sim 1000$~km~s$^{-1}$ velocities.
Current observations show that at least
some magnetar candidates have the velocities of less than
500~km~s$^{-1}$ (see $\S$~\ref{intro}) whereas \citet{dun92} claimed that
magnetars will have large
kick velocities up to $\sim 1000$~km~s$^{-1}$. It is notable that our
results also show that the some moderate initial conditions lead to the
velocities less than $\sim~500$~km~s$^{-1}$ which is not inconsistent
with the observations. On the other hand, if the association between SGR
0526-66 and N49 or SGR 
1627-41 and G337.0-0.1 is true, these magnetar candidates should have
an extremely large kick velocities up to $\sim 1000$~km~s$^{-1}$. In the
context of our numerical simulations, these extraordinarily high
velocities are interpreted as the products of the extreme initial
condition, i.e. very strong and highly asymmetric
magnetic fields as in model SR10. 

Some observations of recent date imply
that some magnetars are originate from massive progenitor 
($\gtrsim 20-50M_{\odot}$) \citep[e.g.][]{gae05,fig05,mun06}. We use a $15
M_\odot$ stellar model although our computation 
involves a formation of magnetars. This is because a $15 M_\odot$ star
is a canonical progenitor of core-collapse computations and it has not
known yet whether a massive-star-origin magnetar is common. However, it
is worth investigating in the future whether the similar results to us are
yielded by the numerical simulation with more massive progenitors. When
the massive progenitors are chosen, it is more natural for us to assume rapid
rotation since \citet{heg05} argued from their calculation
of stellar evolution that more massive stars tend to rotate more
rapidly. According to them, $35 M_\odot$ star will reach the rotation
period of the neutron star of $4.4$ ms in there standard model, though
the magnetic flux of the star is far weaker than that of magnetars and
again this result is not conclusive.

The other kick mechanisms introduced in \S~\ref{intro} does not work in
the situation considered here. For the hydrodynamically-driven and
neutrino-magnetic-field driven mechanism, this is because the kicks are
accompanied with delayed explosion while, in our simulation, prompt
explosion occurs with the help of strong magnetic field and rotation. 
In an electromagnetically-driven mechanism, the acceleration time
scale is $\sim $ 4 s if the field strength on the surface of the
(proto-) magnetar is $\sim 10^{16}$~G, which is much longer than that of
``\textit{magnetohydrodynamically-driven kick}''. Our computation shows
that the initially rapid rotation becomes quite slow (more than 10~ms
rotation period) in $\lesssim~100$~ms after bounce (see
\S~\ref{magrot}). Noted that this mechanism requires the rotation period
of $\sim 1$~ms, the proto-magnetar cannot be substantially accelerated.  

The above discussion is valid only when the initial conditions which we
assume, strong magnetic fields, rapid rotation, and offset dipole fields, are
achieved. Although validity of these assumptions is discussed in
$\S$~\ref{magrot}, they are far from solid. What will happen
if our assumptions are partially fulfilled? With strong magnetic fields
but a slow rotation, the magnetorotational explosion considered here can
not occur. In this case, the engine of supernova might be the heating by
emitted neutrinos, and the second kick mechanism (neutrino-magnetic
field driven) introduced in $\S$~\ref{intro} might work. We
have no idea at the moment whether the hydrodynamical-instability-driven
kick (the first mechanism) could operate or not in the presence of
strong magnetic fields. This should be investigated in the future
works. If the initial magnetic field is weak and rotation is rapid,
 the magnetorotational explosion may ensue with a help of
 magnetorotational instability (MRI) or convective dynamo action
 \citep[e.g.][]{aki03, ard05, dun92}. However, it needs more detailed
 studies. Note that in order to reliably compute the growth of MRI from weak
 magnetic fields of $\sim~10^8-10^9$~G in the supernova core, the spatial grid
 size should be at least as small as $\sim~10^2$~cm, which no numerical
 simulation has achieved yet so far. Meanwhile, even in the present
 computations with $\sim~10^{12}-10^{13}$~G initially, the grid resolution is
 not fine enough by a factor of ten in some important regions to resolve MRI. 
 In this case, however, the compression and wrapping-up of the initial field
 amplify the field strength sufficiently before MRI sets in and we expect MRI
 will not change the dynamics very much even if it occurs.
 But, we do not intend to claim that MRI is not important. On the contrary,
 we are interested in the MRI-related issues such as the linear growth,
 nonlinear saturation, implications for explosion, pulsar kick and spin and so
 forth. We think they should be addressed in a separate paper. In so doing,
 three-dimensional computations are important, which we are now undertaking
 (Sawai et al. 2008 in preparation).  

We find that the centrally concentrated magnetic field has an advantage in
energetic explosion. Recently, \citet{pia06} reported that SN 2006aj
associated with X-ray flash 060218 is a factor 2-3 more luminous than
other normal type Ic supernovae although it is dimmer than so-called
hypernovae associated with gamma-ray bursts. \citet{maz06} claimed that
the remnant of SN 2006aj is not a black hole but a neutron star,
possibly a magnetar. If this is true, a highly
energetic explosion induced by centrally-concentrated magnetic fields can
be one of the origins of this middle-class supernova although the
mechanism to produce the X-ray flash is beyond the scope of this
paper. Note that not every supernova accompanied by a magnetar
should produce such a energetic explosion. It is reported that
for some supernovae, which are considered to be associated with 
magnetar candidates, the explosion
energies are just as large as those of ordinary supernovae
\citep{vin05,sas04}.

Finally, how large a fraction do magnetars occupy among the whole population of
isolated neutron stars? At present, the confirmed candidates of magnetars are
counted up to eleven which is only $\sim 1$ \% of the observed
ordinary pulsars. However it is too early to conclude that magnetars
are such a minor population. Spin-down ages are
measured in three of four SGRs, which are younger than 1900 years. This means
a magnetar is born every $\sim 600$ years, corresponding to $\sim 15$ \% of
ordinary neutron stars. If another magnetar, SGR 1627-41, has a similar
spindown age, the number of magnetars are estimated to be $\sim
20$ \% of that of the ordinary pulsars. Here we omit AXPs since there
may be a large number of AXPs which are too dim to detect
\citep{woo06}. \citet{woo06} insisted that there
may be a lot of magnetars which are too old to be detected
and that the number of magnetars could be comparable to that of ordinary
pulsars. If this is the case, the study of the supernova theory in the
strong-magnetic-field regime will have a more significant meaning.
 
\acknowledgments
H. S. thanks to E. M\"{u}ller and H.-Th. Janka for useful discussion during his
stay at Max-Planck-Institute f\"{u}r Astrophysik. H. S. also thanks
D. Lai for helpful discussion.     
Some of the numerical simulations were done on the
supercomputer VPP700E/128 at RIKEN and VPP500/80 at KEK (KEK
Supercomputer Projects No. 108). This work was partially supported by
the Japan Society for Promotion of Science (JSPS) Research Fellowships
(H.S.),
the Grants-in-Aid for the Scientific 
Research (14079202, 17540267) from Ministry of Education, Science and
Culture of Japan, and by the Grants-in-Aid for the 21th century COE
program ``Holistic Research and Education Center for Physics of
Self-organizing Systems''.

\begin{figure}
\plotone{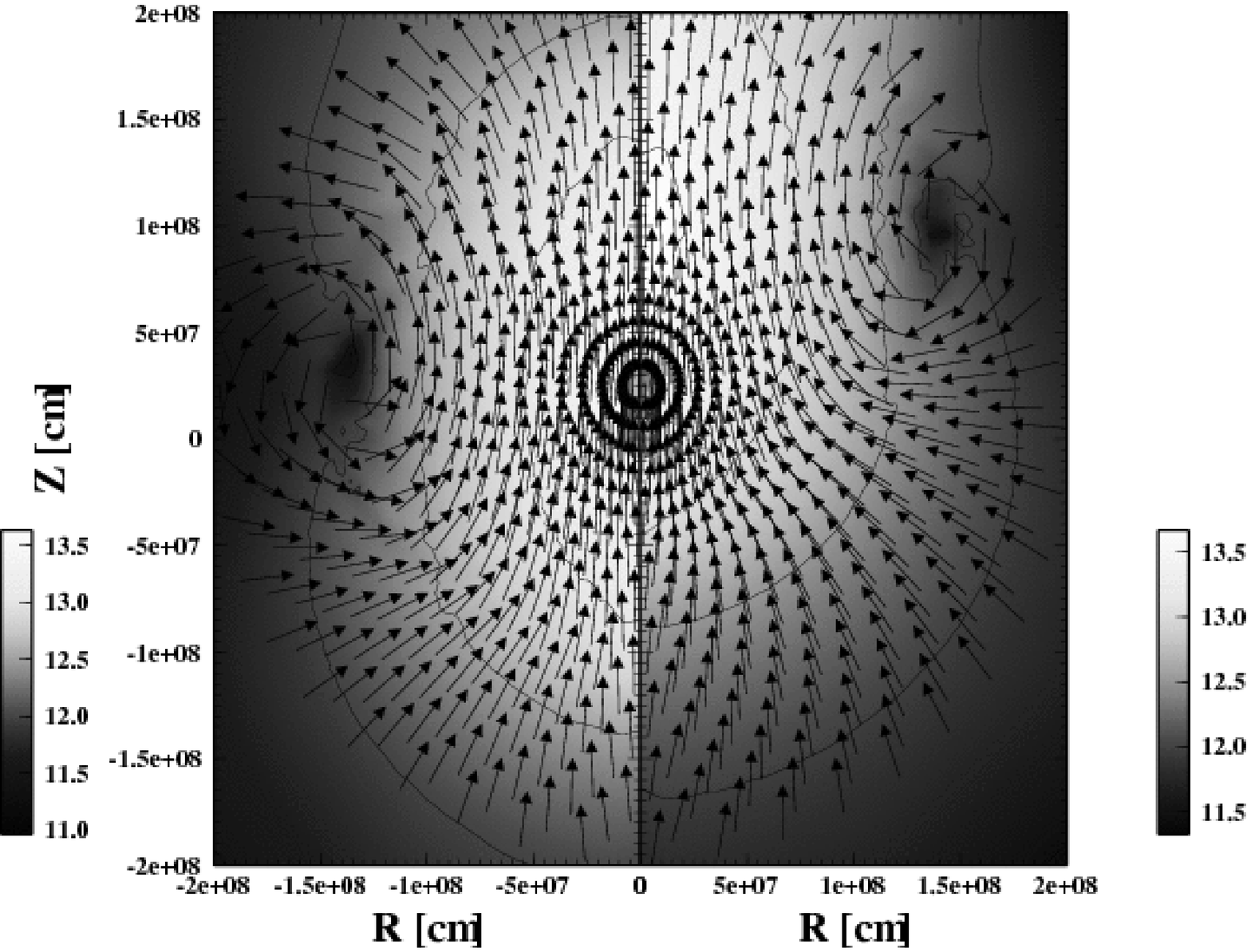}
  \caption{The initial magnetic magnetic fields of models MR3 (left
 panel) and SR10 (right panel). Gray colors show the strengths of
 the poloidal magnetic fields and the vectors represent the directions of
 the magnetic fields. The initial radius of the 1.5
 M$_{\odot}$ core is 2000~km.} 
 \label{fbfield}
\end{figure}

\begin{figure}
\epsscale{1}
\plotone{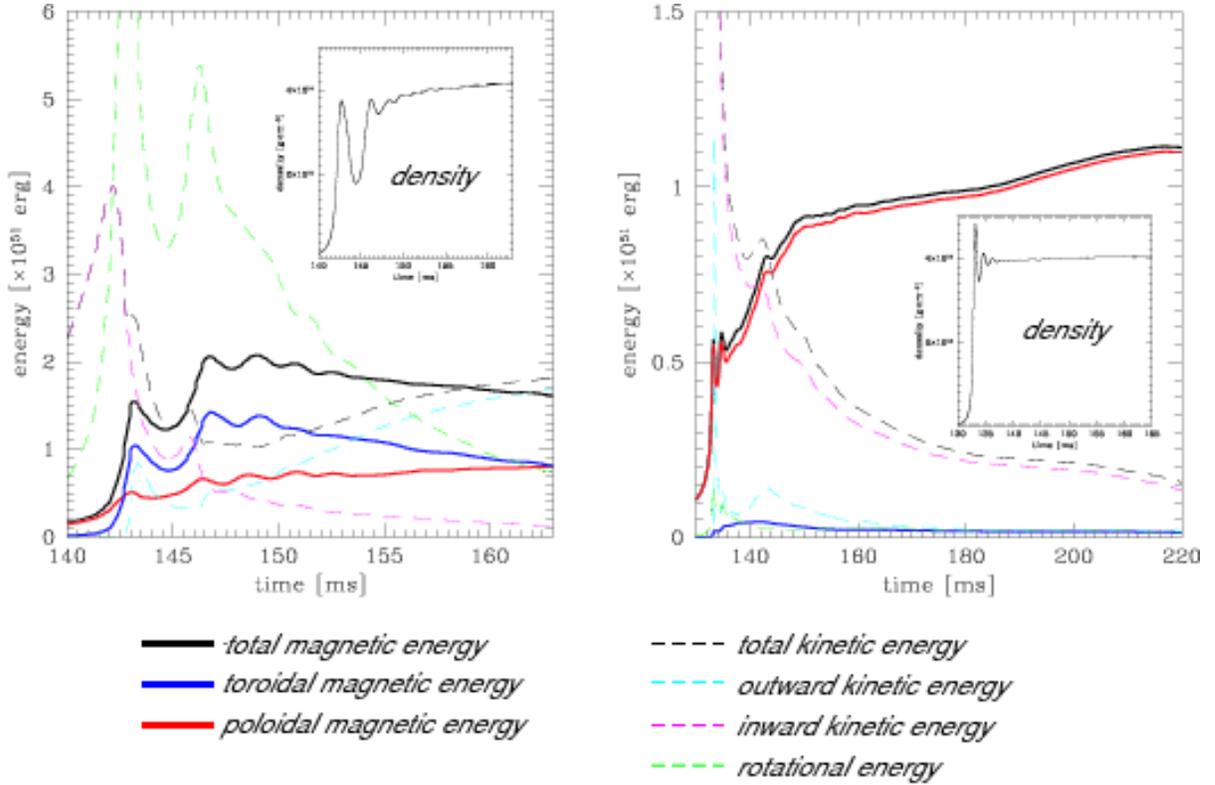}
  \caption{The time evolutions of the rotational, kinetic, poloidal magnetic,
 and toroidal magnetic energy until the farthest shock reaches a
 radius 1500~km. the left and right panels correspond to model MR0 and
 MS3, respectively. ``Outward kinetic energy'' and ``inward kinetic
 energy'' are defined as the kinetic energy of matters whose radial
 velocity is positive and negative, respectively. } 
 \label{fengy}
\end{figure}

\begin{figure}
\plotone{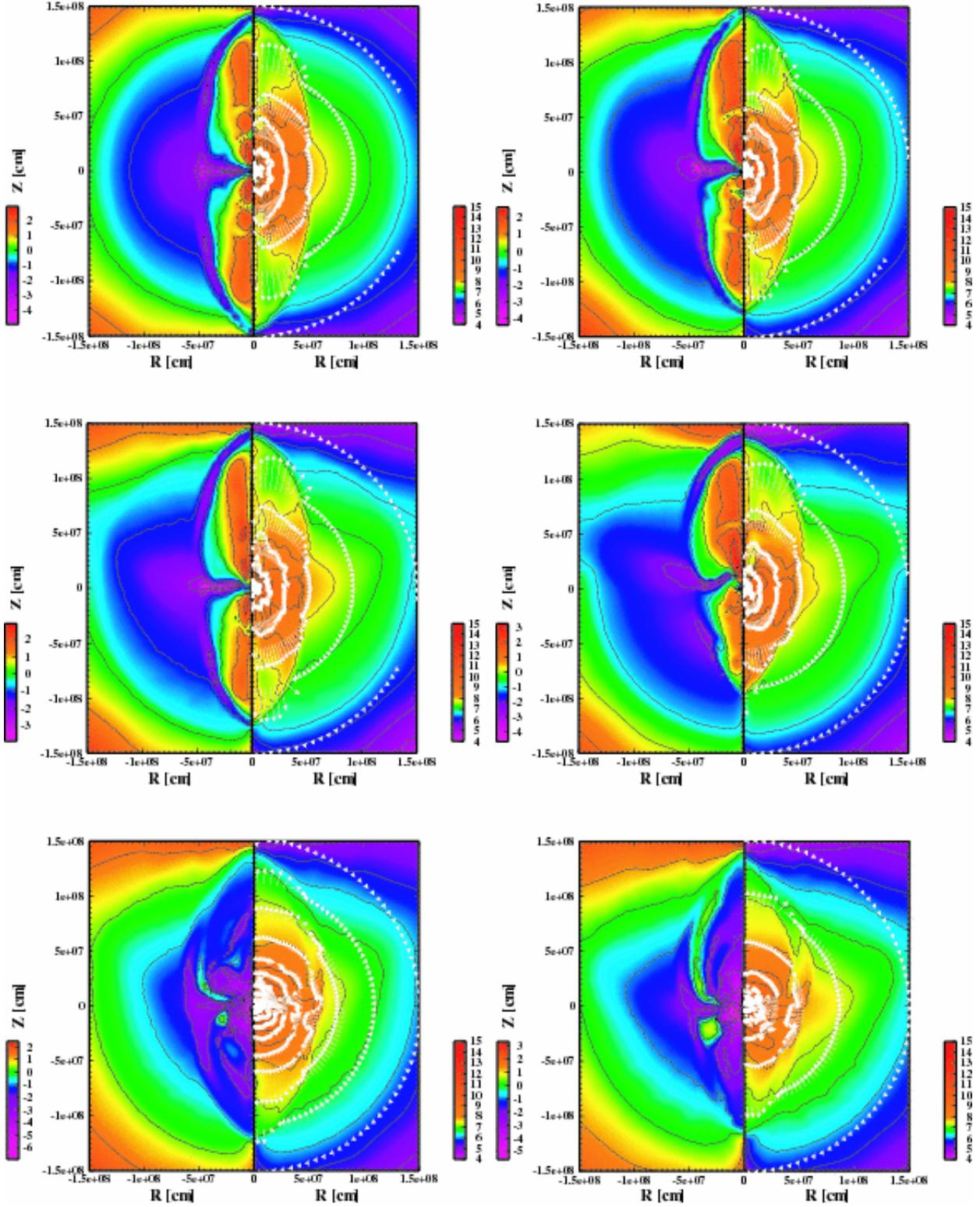}
  \caption{The velocity fields on top of the density color contours (the
 right half of each panel), and the contours of the ratio of the
 magnetic pressure to matter pressure (the left half of each
 panel) when the shock front
 reaches a radius of about 1500~km. From left to right
 and top to bottom, panels of model MR0, MR3, SR3,
 SR10, MS3, and SS3 are displayed in sequence. In each panel, colors in
 the region out of a radius of 1500~km are extrapolated.}  
 \label{fvbeta}
\end{figure}


\begin{figure}
\epsscale{1}
\plotone{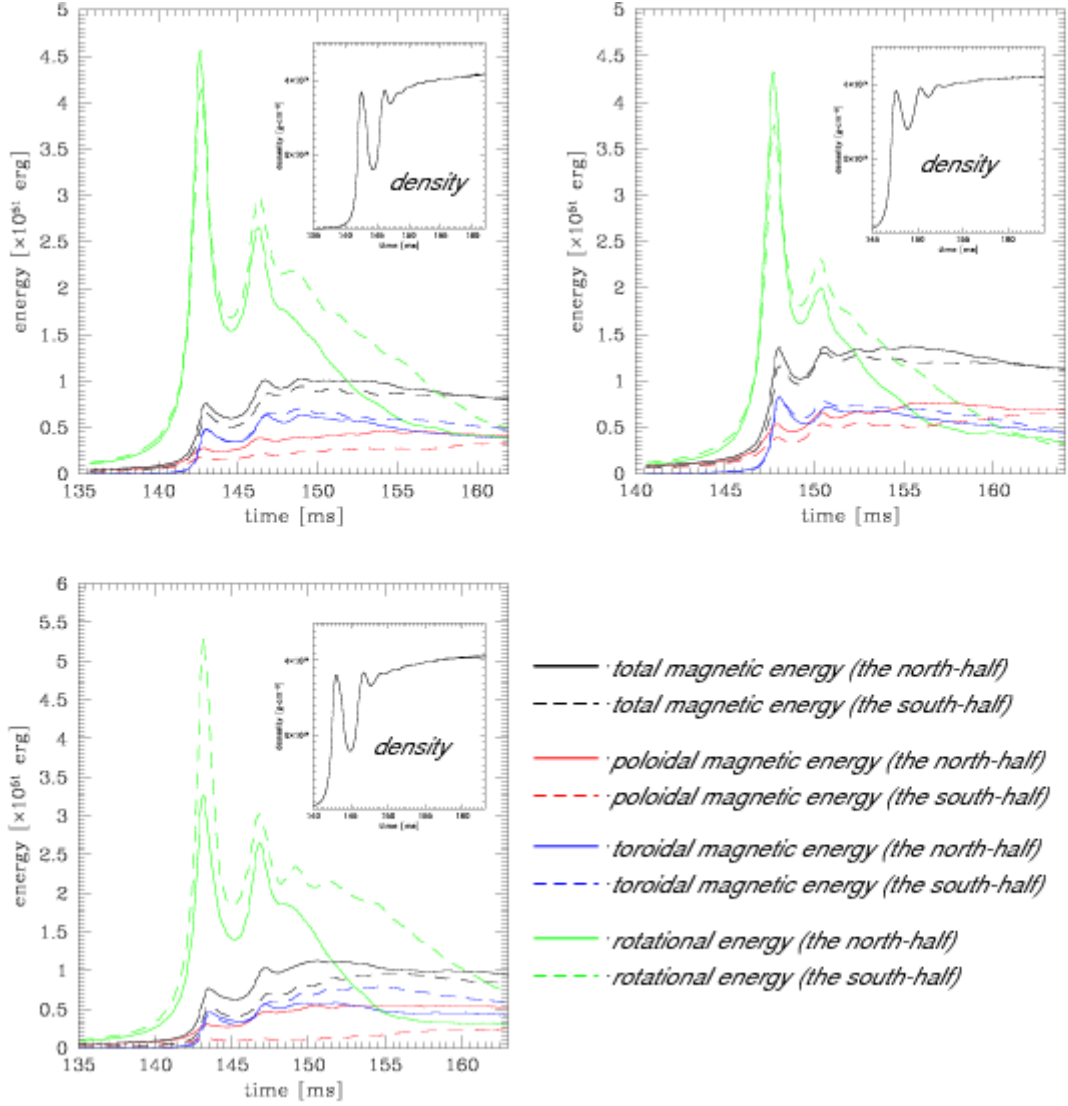}
  \caption{The time evolution of the total, poloidal, and toroidal magnetic
 energy and the rotational energy. The energies of the northern and
 southern hemisphere are drawn separately. The insets show the time evolution of the
 central density. Figures of models MR3, SR3, and SR10 are put in order
 from left to right and top to bottom.} 
 \label{fengmag}
\end{figure}

\begin{figure}
\plotone{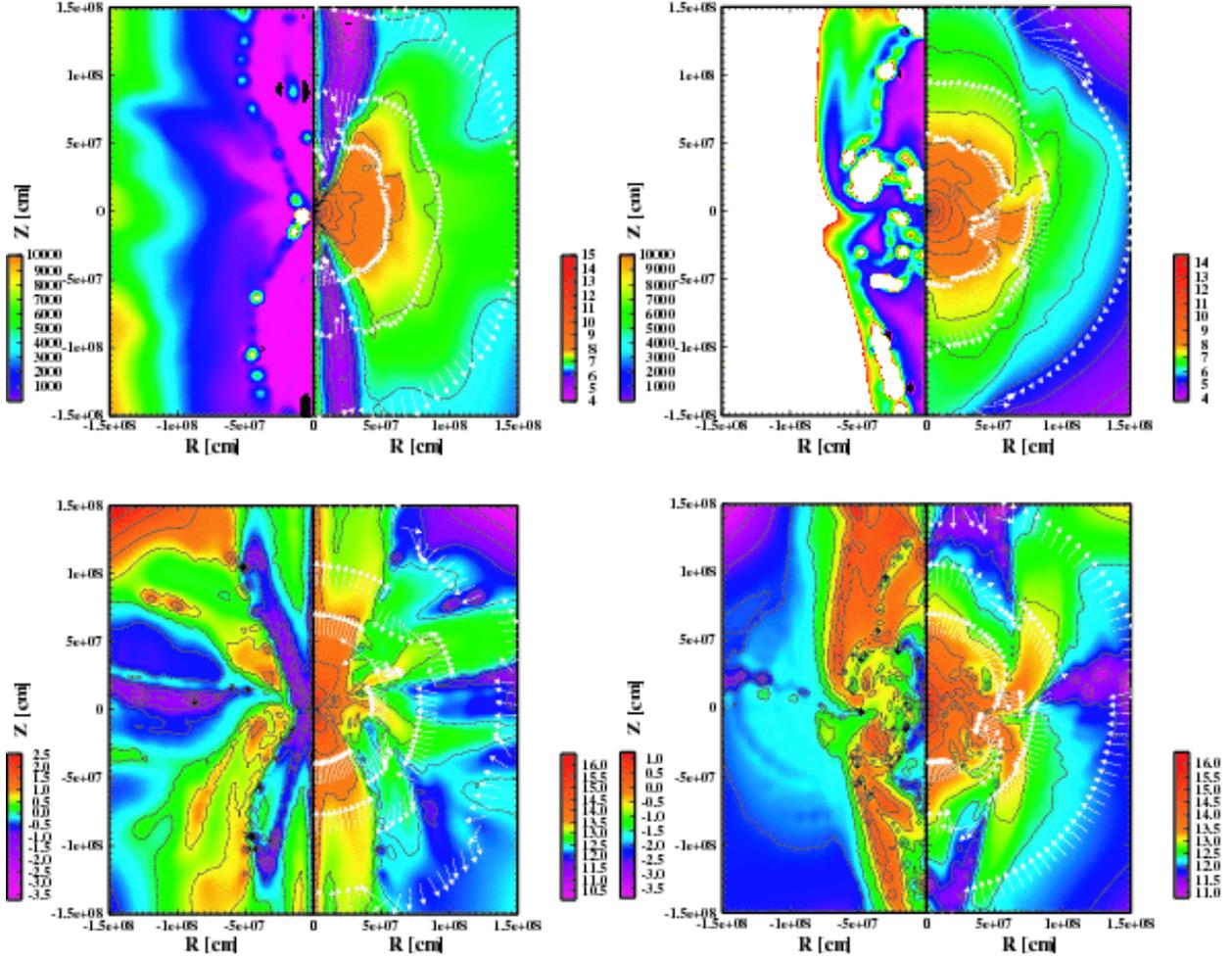}
  \caption{Upper two panels: The density contours (the right half of
 each panel), and the contours of the rotation period in millisecond
 (the left half of each panel) at 240 ms from the beginning for models
 MR3 (the left panel) and MS3 (the right panel). In the left half of
 each panel, the white-colored and black-colored regions have rotation period
 of more than 10 s and less than 10 ms, respectively. The vectors in the
 right-half of each panel represent the velocity field. Lower two panels:
 The contours of the poloidal field strength (the right half of
 each panel), and the contours for the ratio of the toroidal to poloidal field
 strength (the left half of each panel) at 240 ms from the beginning for models
 MR3 (the left panel) and MS3 (the right panel). The vectors in the
 right-half of each panel represent the directions of the poloidal
 magnetic field.} 
 \label{fmagrot}
\end{figure}

\begin{figure}
\plotone{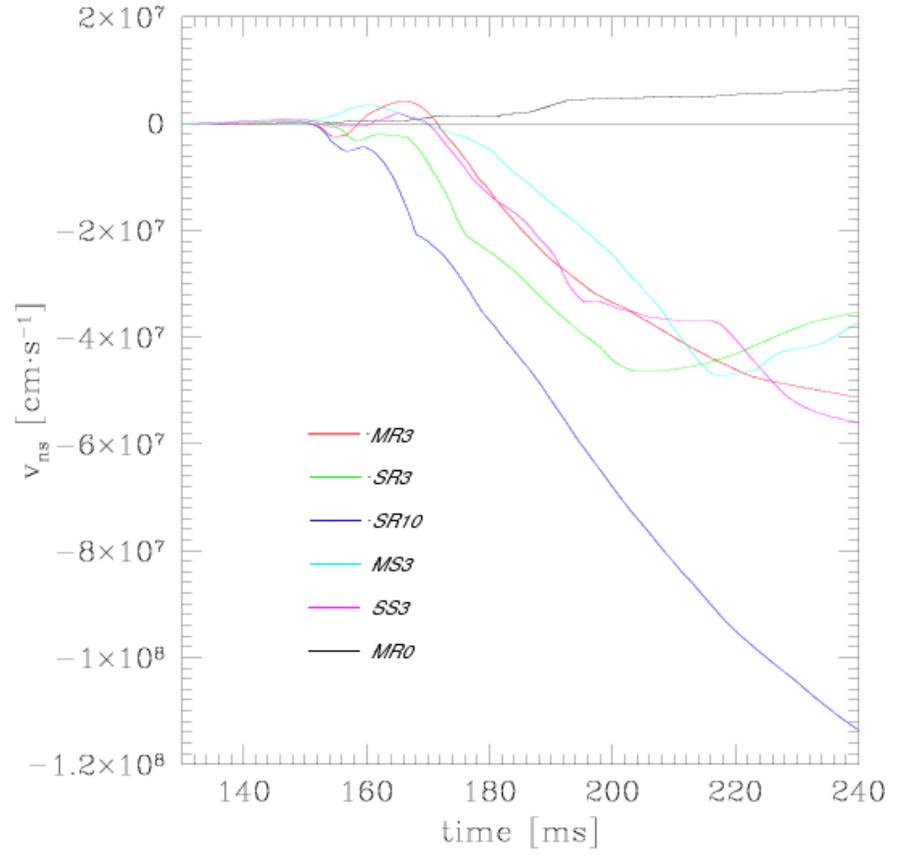}
  \caption{Time evolutions of the proto-magnetar velocities. The northward
 velocity is taken to be positive.} 
 \label{fvns}
\end{figure}

\begin{figure}
\begin{center}
\plotone{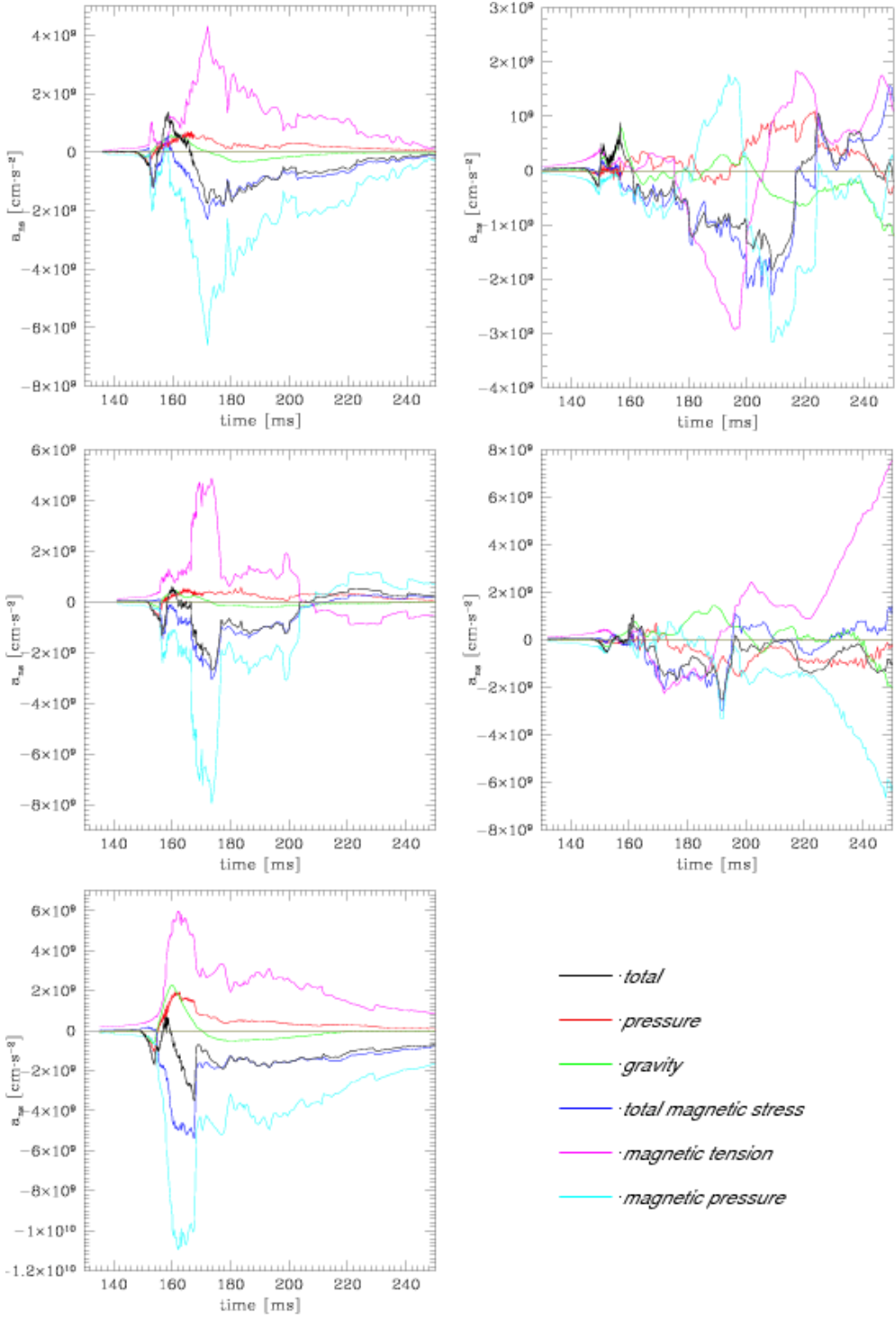}
  \caption{Time evolutions of the accelerations acting on the proto-magnetar
 star. The magnetic acceleration divided into two parts, the
 tension part and pressure part. From top
 to bottom and left to right, panels of model MR3, SR3,
 SR10, MS3, and SS3 are displayed in sequence. The northward
 acceleration is taken to be positive.} 
 \label{fans}
\end{center}
\end{figure}

\begin{figure}
\begin{center}
\plotone{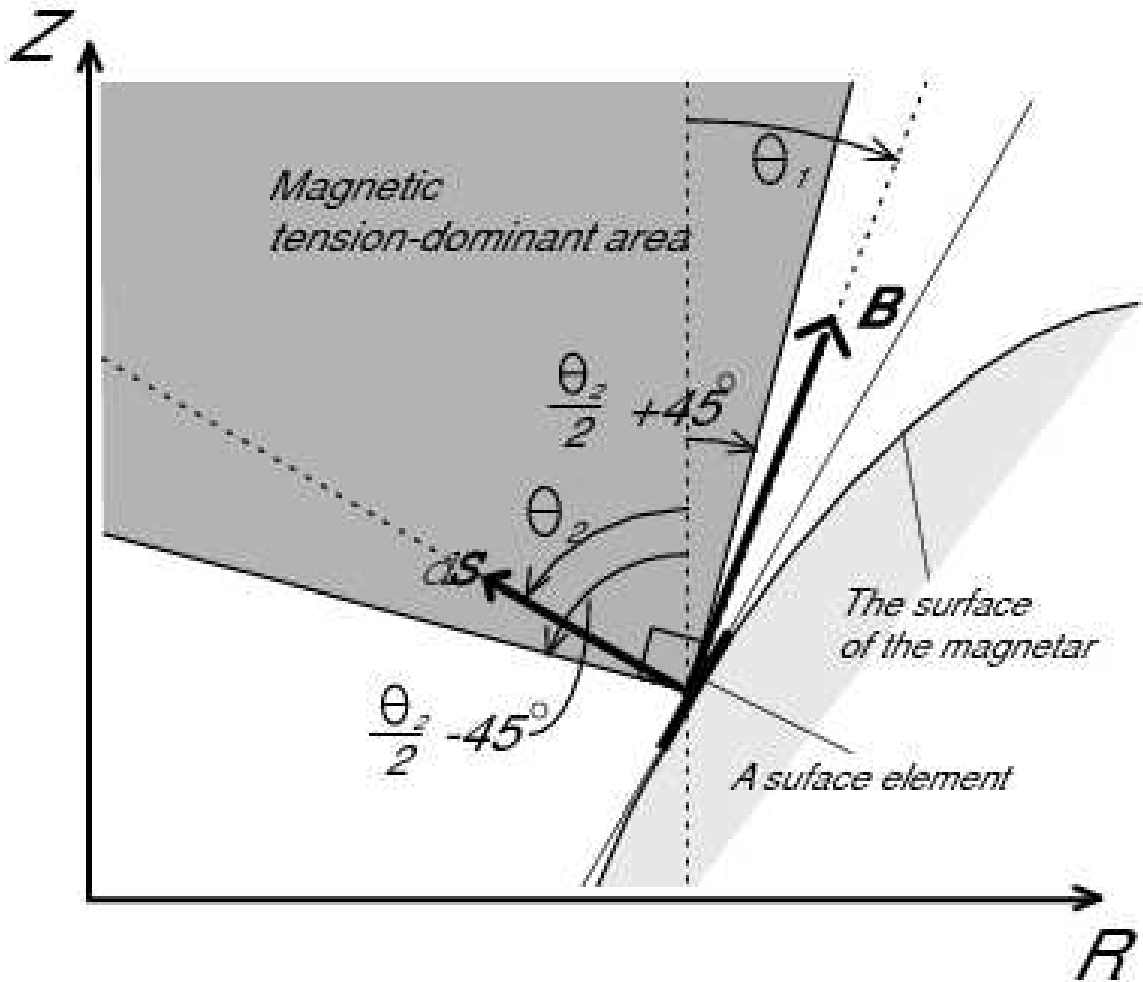}
  \caption{A schematic picture which shows the magnetic tension-dominant
 area (strong gray colored). \mbox{\boldmath $B$} and $d$\mbox{\boldmath $S$}
 is the magnetic field vector and surface vector, respectively. $\theta_1$ and
 $theta_2$ are the angles from the
 rotation axis to the magnetic field vector and surface vector, respectively,
 taking clockwise rotation positive.}
 \label{fmagtp}
\end{center}
\end{figure}

\begin{figure}
\plotone{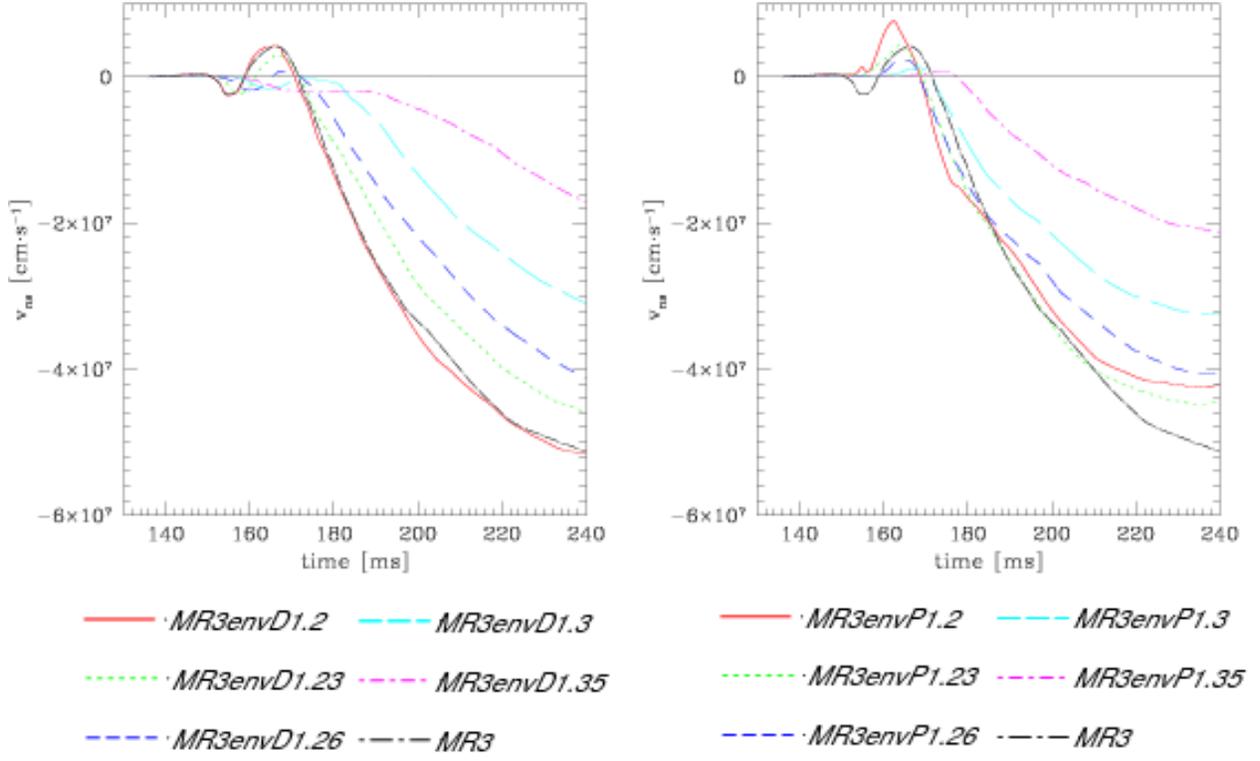}
  \caption{Time evolutions of the proto-magnetar velocities for several
 definitions of a proto-magnetar. Left and right panel corresponds to the
 results 
 of model series of MR3evnD and MR3envP, respectively. In each panel, solid,
 dotted, short-dashed, long-dashed, and dotted-short-dashed, 
 lines correspond to model of 1.2, 1.23, 1.26, 1.3, and 1.35~M$_\odot$
 proto-magnetars, whereas dotted-long-dashed line is drawn for model MR3 for
 comparison. The northward
 velocity is taken to be positive.} 
 \label{fvnsmr3env}
\end{figure}

\begin{table}
\begin{center}
\caption{Initial Parameters}\label{tmodels}
\begin{tabular}{lccccc}
\tableline\tableline
Model & $|E_m/W|$ [\%] & $|T/W|$ [\%]& $B_i$ [G] & $\Omega_i$ [rad s$^{-1}$] &
 $z_{off}$ [km]\\
\tableline
MR0  & 0.5 & 0.5   & $5.6 \times 10^{13}$  & $3.9 \times 10^{0}$ & 0 \\
MR3  & 0.5 & 0.5   & $4.6 \times 10^{13}$  & $3.9 \times 10^{0}$ & 300 \\
MS3  & 0.5 & 0.005 & $4.6 \times 10^{13}$  & $3.9 \times 10^{-1}$ & 300 \\
SR3  & 1.0 & 0.5   & $6.5 \times 10^{13}$  & $3.9 \times 10^{0}$ & 300 \\
SR10 & 1.0 & 0.5   & $4.9 \times 10^{13}$  & $3.9 \times 10^{0}$ & 1000 \\
SS3  & 1.0 & 0.005 & $6.5 \times 10^{13}$  & $3.9 \times 10^{-1}$ & 300 \\

\tableline
\end{tabular}
\tablecomments{$|E_m/W|$: the magnetic energy normalized by the
 gravitational energy. $|T/W|$: the rotation energy normalized by the
 gravitational energy. $B_i$: the initial maximum magnetic
 field. $\Omega_i$: the initial angular velocity at the center of
 the core. $z_{off}$: the degree of magnetic-field displacement (see
 Eq.~\ref{emag1}-~\ref{emag2}).}
\end{center}
\end{table}

\begin{deluxetable}{lcccccccccccc}
\tabletypesize{\footnotesize}
\rotate
\tablecaption{Key Parameters}\label{tpara}
\tablewidth{0pt}
\tablehead{
\colhead{Model} & 
 \colhead{$|E_m/W|_b$} &\colhead{$|T/W|_b$} 
&\colhead{$B_b,max$}   &\colhead{$\Omega_b,max$} 
&\colhead{$|E_m/W|_{t_1}$} &\colhead{$|T/W|_{t_1}$} 
&\colhead{$B_{t_1,max}$}   &\colhead{$\Omega_{t_1,max}$} 
&\colhead{$r_{sh,u}/r_{sh,d}$}
& \colhead{$v_{NS}$}}

\startdata
MR0 & 1.5   & 8.0   & $5.3 \times 10^{16}$ & $6.5 \times 10^{4}$ & 
      1.6  & 0.71 & $6.3 \times 10^{16}$ & $1.6 \times 10^{4}$ &
      0.97 & 65\\ 
MR3 & 1.4   & 8.2   & $8.3 \times 10^{16}$ & $7.3 \times 10^{4}$ & 
      1.6  & 0.87 & $6.2 \times 10^{16}$ & $1.9 \times 10^{4}$ &
      1.15 & 512\\ 
MS3 & 0.50  & 0.13  & $1.2 \times 10^{17}$ & $5.8 \times 10^{3}$ & 
      0.98  & 0.016 & $3.1 \times 10^{16}$ & $1.6 \times 10^{2}$ &
      1.20 & 374\\ 
SR3 & 1.6  & 7.5  & $1.1 \times 10^{17}$ & $6.0 \times 10^{4}$ & 
      2.1  & 0.59 & $8.1 \times 10^{16}$ & $2.6 \times 10^{4}$ &
      1.17  & 353\\ 
SR10& 0.76  & 7.8  & $9.0 \times 10^{16}$ & $7.9 \times 10^{4}$ & 
      1.6  & 1.1 & $4.8 \times 10^{16}$ & $1.4 \times 10^{4}$ &
      1.51  & 1136\\
SS3 & 0.94  & 0.089  & $1.3 \times 10^{17}$ & $2.1 \times 10^{4}$ & 
      1.5   & 0.021 & $4.7 \times 10^{16}$ & $3.6 \times 10^{4}$ &
      1.21 & 560\\  
\enddata

\tablecomments{The ratios $|E_m/W|$ and $|T/W|$ are given in
 percentage. $B_{max}$ : the maximum magnetic field in
 G. $\Omega_{max}$ : the maximum angular velocity in rad s$^{-1}$. 
 $r_{sh,u}/r_{sh,d}$: the ratio north-shock radius to
 south-shock radius. $v_{NS}$: the velocity of the magnetar in km
 s$^{-1}$ at 240 ms from the beginning. 
 For each parameter, the subscripts ``$b$'' and ``$t_1$''
 denote the values at bounce and those when the shock front reaches
 radius of 1500 km, respectively.}
\end{deluxetable}
\begin{table}
\begin{center}
\caption{Ejected Mass and Explosion Energy}\label{texp}
\begin{tabular}{lcccccc}
\tableline\tableline
Model & $M_{ej,tot}$ & $M_{ej,n}$ & $M_{ej,s}$ & $E_{exp,tot}$ & $E_{exp,n}$ & $E_{exp,s}$ \\ 

\tableline
MR0  & 0.31  & 0.15  & 0.15  & 4.4  & 2.2  & 2.2  \\
MR3  & 0.28  & 0.11  & 0.16  & 4.1  & 1.8  & 2.3  \\ 
MS3  & 0.029 & 0.017 & 0.012 & 0.38 & 0.20 & 0.17 \\
SR3  & 0.25  & 0.11  & 0.15  & 4.6  & 2.1  & 2.5  \\
SR10 & 0.27  & 0.095 & 0.17  & 4.0  & 1.8  & 2.4  \\
SS3  & 0.036 & 0.021 & 0.015 & 0.56 & 0.33 & 0.23 \\

\tableline
\end{tabular}
\tablecomments{$M_{ej}$ : the ejected mass in $M_\odot$. $E_{exp}$ : the
 explosion energy in $10^{51}$ erg. Subscripts ``n'', ``s'', and ``tot''
 denote the values in the northern hemisphere, the southern
 hemisphere, and total values, respectively.}
\end{center}
\end{table}

\end{document}